\documentclass[conference]{IEEEtran}
\IEEEoverridecommandlockouts
\usepackage{cite}
\usepackage{amsmath,amssymb,amsfonts}
\usepackage{algorithmic}
\usepackage{graphicx}
\usepackage{textcomp}
\usepackage{soul}
\usepackage{caption}
\usepackage{subcaption}
\usepackage{tikz}
\usepackage{booktabs}
\usepackage{enumitem}
\usepackage{xcolor}
\def\BibTeX{{\rm B\kern-.05em{\sc i\kern-.025em b}\kern-.08em
    T\kern-.1667em\lower.7ex\hbox{E}\kern-.125emX}}
\begin{document}

\title{Cooperative Driving for Speed Harmonization in Mixed-Traffic Environments}

\author{Zhe Fu$^{1,2}$, Abdul Rahman Kreidieh$^{2}$, Han Wang$^{2}$\\ 
Jonathan W. Lee$^{2}$, {Maria Laura} {Delle Monache}$^{2}$ and Alexandre M. Bayen$^{2}$
\thanks{$^{1}$Corresponding author, email: zhefu@berkeley.edu}%
\thanks{$^{2}$University of California, Berkeley}%
}
\maketitle

\begin{abstract}
Autonomous driving systems present promising methods for congestion mitigation in mixed autonomy traffic control settings. In particular, when coupled with even modest traffic state estimates, such systems can plan and coordinate the behaviors of \emph{automated vehicles} (AVs) in response to observed downstream events, thereby inhibiting the continued propagation of congestion. 
In this paper, we present a two-layer control strategy in which the upper layer proposes the desired speeds that predictively react to the downstream state of traffic, and the lower layer maintains safe and reasonable headways with leading vehicles. This method is demonstrated to achieve an average of over $15\%$ energy savings within simulations of congested events observed in Interstate 24 with only $4\%$ AV penetration, while restricting negative externalities imposed on traveling times and mobility. The proposed strategy that served as part of the ``speed planner" was deployed on 100 AVs in a massive traffic experiment conducted on Nashville's I-24 in November 2022.
\end{abstract}


\section{Introduction}
Vehicle autonomy is rapidly becoming a viable feature of many road networks. Early demonstrations in vehicle platooning~\cite{shladover2007path, robinson2010operating, tsugawa2011automated} and similar successes spurred on by ambitious driving challenges~\cite{buehler20072005, buehler2009darpa} have motivated equally ambitious efforts 
in the industrial sector, with companies including Tesla~\cite{tesla}, Google~\cite{waymo}, GM~\cite{cruise}, and others all attempting to push the limitations and scope of vehicle autonomy. 
This trajectory is expected to continue as well, with studies projecting and discussing the implications of autonomy in the vicinity of $20$-$40$\% by $2050$~\cite{litman2017autonomous}.


In this paper, we are interested primarily in the role of \emph{longitudinal} driving behaviors on the energy-efficiency of a given network. To improve energy-efficiency, we need to dampen traffic oscillation, and one well-known approach for this is speed harmonization which aims to reduce temporal and spatial variations of traffic speed by applying certain control approaches~\cite{ma2016freeway}. This is a topic that has formerly been heavily explored in the context of platoons of connected and automated vehicles(CAVs), whereby platoons of fully-automated vehicles have successfully maintained string-stable driving responses in tight platoons, thereby providing notable benefits to both energy-efficiency and throughput.
More relevant to the present paper, however, AVs in mixed-autonomy settings may provide significant benefits in mitigating string-instabilities among \emph{human drivers} as well. 
In the work~\cite{ghiasi2019mixed}, a mixed traffic speed harmonization algorithm that is based on trajectory prediction from sensor data and probe data was proposed to control CAVS so that they smoothly hedge against the backward deceleration waves and gradually merge into the downstream traffic with a reasonable speed.
It was also demonstrated empirically in the seminal work of~\cite{stern2018dissipation}, whereby a single AV within a circular track stably operating near the effectively uniform driving speed of the network manages to dampen stop-and-go oscillations existing prior to actuation.

The above empirical study provides a useful insight that is frequently mentioned~\cite{cui2017stabilizing,kreidieh2022learning}: significant gains to energy-efficiency may be achieved by harmonizing the speeds of subsets of vehicles near a desirable target. This deduction, however, introduces new challenges to autonomous driving systems. In particular, under the ever-evolving dynamics of a particular network as demand waxes and wanes, AVs must reactively identify desirable speeds that match current spatio-temporal trends while not inhibiting the safety or mobility of the vehicle. To this, traffic state estimates may offer a helping hand. Estimates of flow, density, and speed produced either from fixed sensors or probe vehicles~\cite{seo2017traffic} may elucidate spatio-temporal patterns that may be exploited by AVs in a largely decentralized manner. This is in part demonstrated in the work of~\cite{asadi2010role}, for instance, an optimal speed profile for vehicles can be generated with devices to provide speed measurements forward in space and time. Solutions such as these, however, are often studied in the context of a fully-observable macroscopic environment, and as such become brittle and unsafe in the presence of inaccurate traffic state estimates and microscopic fluctuations in speed and spacing.



In this paper, we present a two-layer control strategy that exploits both macroscopic traffic state estimates(TSE) and microscopic observations to produce a reasonable car-following response while also attempting to harmonize driving speeds across a desirable spatio-temporal target. The key contributions of this paper are as follows:

\begin{itemize}
    \item We construct a two-layer longitudinal feedback control strategy for AVs in which the upper layer generates the desired speed that helps to smooth the traffic flow, and the lower layer attempts to maintain reasonable headways with preceding vehicles when appropriate. 
    \item We validate the efficacy of the above method on a simulation of throughput-restricted traffic aimed at capturing the high degree of variability in driving behaviors 
    and traffic state estimates 
    common to real-world networks, and we demonstrate
    that our method can consistently achieve smoother traffic flow and large energy savings 
    in congested states of traffic.
    \item We justify the modular and flexible architecture of the two-layer control strategy by deploying the upper layer as part of the ``speed planner" for 100 AVs in a massive traffic experiment conducted on Nashville's I-24 in November 2022. 
\end{itemize}


\section{Controller Design}

\subsection{Problem Statement}
\begin{figure}
\centering
\includegraphics[height=4.6cm]{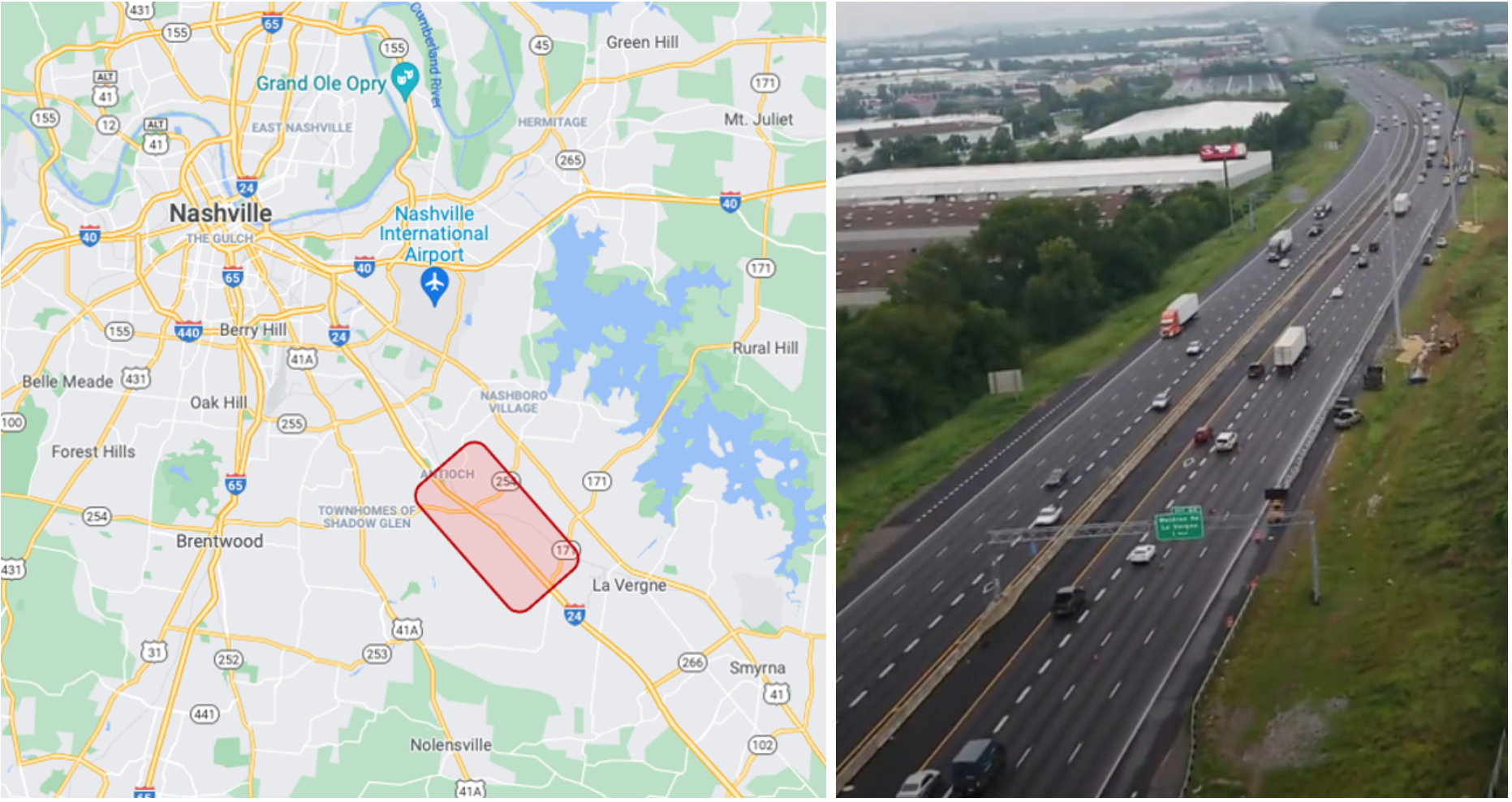}
\caption{An illustration of the targeted highway network within this study (I-24 Westbound in Nashville, Tennessee), seen within the highlighted region.}
\label{fig:network}
\end{figure}

In this paper, we are interested in exploring methods for ameliorating congestion in mixed autonomy highway networks. The considered network, see Figure~\ref{fig:network}, is a 14.5-km long segment of I-24 located in Nashville, Tennessee. This network has been the topic of some interest in recent years, with researchers attempting to both reconstruct~\cite{lee2021integrated,lichtle2022deploying} and address~\cite{kardous2022rigorous, hayat2022holistic, lichtle2022deploying} characteristics of driving within this network that produce inefficiencies in energy consumption. In particular, we explore the implications of automated driving on addressing inefficiencies arising from string instabilities in human driving behaviors, which result in the formation of stop-and-go traffic during peak demand intervals within this network.

\begin{figure}
\centering
\includegraphics[height=2.5cm]{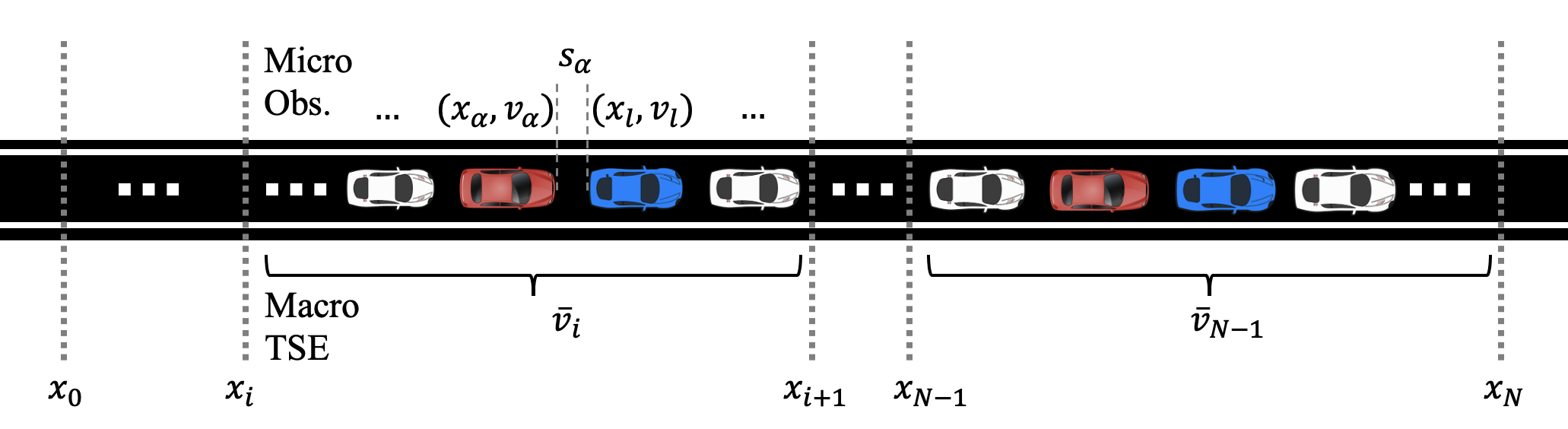}
\caption{A Top View interpretation of the explored single-lane problem setup. Red cars represent AVs, blue represents sensed leading human-driven vehicles and white represents unsensed human-driven vehicles. Grey dashed vertical lines indicate the division of macroscopic TSE segments and their positions.}
\label{fig:problem}
\end{figure}

We focus on deriving longitudinal responses for AVs dispersed with a straight, single-lane track. Figure~\ref{fig:problem} provides a visual interpretation of the explored problem setup. We consider the task of traffic flow harmonization in settings where AVs (in red) are distributed amongst, and interact with, human-driven vehicles (in blue and white, blue represents the leading vehicle sensed by AV). Both microscopic observations and macroscopic traffic state estimations (TSE) can be obtained by AVs. We assume that besides its own position and velocity information $(x_\alpha, v_\alpha)$ the AV can sense the leading vehicle's position and velocity $(x_l, v_l)$, and get access to downstream TSE in the form of space-mean speeds $(x_i, \Bar{v_i})$, $i$ represents the $i_\text{th}$ segement. Based on these inputs, we aim to develop driving speeds for AVs to smooth the traffic flow while maintaining safe and appropriate gaps between AVs and their leaders. 

In the following subsections, we present a two-layer control strategy that includes an upper layer: target speed, a lower layer: gap regulation, and a safety filter.

\subsection{Upper layer: Target Speed}
The proposed approach adopts and extends prior heuristic on traffic flow harmonization~\cite{asadi2010role,cui2017stabilizing,stern2018dissipation,kreidieh2022learning}, which posit that traffic may be homogenized near its desirable \emph{uniform} driving speed by operating a subset of vehicles near accurate predictions of said speed.
The desirable uniform driving speed must be achieved without shared communication between adjacent vehicles, as in mixed-autonomy settings human-driven vehicles are incapable of sharing their desired speeds. Instead, we rely on macroscopic TSE data to synchronize the driving speeds of automated vehicles. In particular, vehicles are assigned target speed profiles contingent on traffic state information, which is shared and common among all AVs. 

The target speed $v_{\text{target}, \alpha} = f(h_\alpha, v_\alpha, v_{\text{des},\alpha})$ is designed based on an actuation function as follows:
\begin{equation}
    v_{\text{target},\alpha} =  \label{eq:actuation}
    \begin{cases}
        v_\alpha & \text{if } \ 0 \leq h_\alpha < 1 s\\
        (2-h_\alpha)v_\alpha + (h_\alpha-1)v_{\text{des},\alpha} & \text{if } \ 1 s \leq h_\alpha \leq 2 s\\
        v_{\text{des},\alpha} & \text{if } \ h_\alpha > 2 s
    \end{cases},
\end{equation}
where $v_\alpha$ and $h_\alpha$ are the speeds and time gaps (in seconds $s$) of the subject vehicle, respectively.

To address safety concerns, we assign the desired speed only if the time gap between the subject and the preceding vehicle is not relatively small. This design is consistent with car-following behavior, which prioritizes maintaining a safe distance between vehicles.

The desired speed profile $v_{\text{des}, \alpha}$ may be extracted in a number of different ways, with many convolutional mappings capable of homogenizing the flow of traffic. Given that macroscopic TSE data is typically quite sparse in real-world applications, such as the INRIX data~\cite{cookson2017inrix} available for our considered highway network has a granularity of approximately one average speed data point for every half-mile, we utilized kernel methods to obtain the desired speed. First, we preprocess the sparse TSE data by interpolating those discrete data pairs $(x_i, \Bar{v_i})$ to a continuous speed profile $v_j = v(x_j)$, $j$ represents any points on the road, as an approximation of the traffic state with higher granularity. Then we obtain the desired speed by applying a kernel function $K(\cdot)$ on $v(x_j)$ given the position of the AV $x_\alpha$:
\begin{equation}
    v_{\text{des},\alpha} = \frac{\int_{x=x_\alpha}^{x_\alpha+w} K(x_\alpha, x) v(x)dx}{\int_{x=x_\alpha}^{x_\alpha+w} K(x_\alpha, x)dx}, \label{eq:desire_kernel}
\end{equation}
where $x_\alpha$ is the position of the subject vehicle, and $w$ is the width of the estimation window.

Many different kernel functions such as Gaussian kernel, Triangular kernel, Quartic kernel, Uniform kernel, etc. can be chosen.
For the purposes of this study, we consider a uniform kernel, the simplest such mapping. The desired speed profile at a position $x_\alpha$ is accordingly defined as:
\begin{equation}
    v_\text{des}(x_\alpha) = \frac{\int_{x=x_\alpha}^{x_\alpha+w} v(x)dx}{w}. \label{eq:desire}
\end{equation}
 
\begin{figure}
    \centering
    \includegraphics[width=\linewidth]{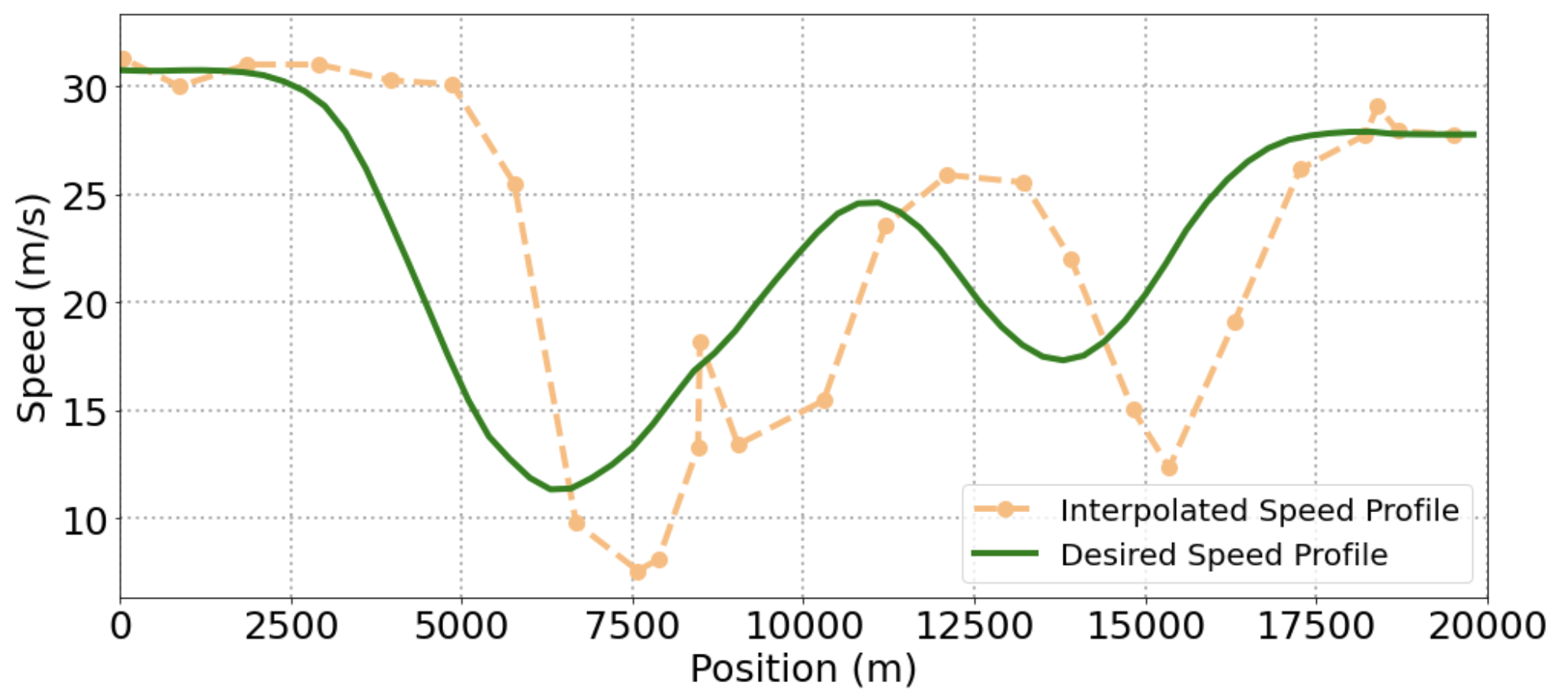}
    \caption{Example of the proposed desired speed profile for sample traffic state estimates. Average (aggregated) speeds across multiple segments (orange dots) are interpolated to a continuous profile. Uniform kernel as Eq.~\eqref{eq:desire} expressed is applied to obtain the desired speed profile (green line).}
    \label{fig:desire-speed}
\end{figure}

For human drivers, when they observe a gap between their vehicle and the one preceding, they tend to accelerate to close the distance. Ironically, this behavior by human drivers is responsible for causing string instabilities, which lead to traffic congestion and delays. As demonstrated in Figure~\ref{fig:desire-speed}, our proposed desired speed profile aims to slow down \emph{in advance}, although not excessively, to create a gap from the preceding vehicle. This approach takes into account the information provided by the TSE, which indicates the presence of congestion in the nearby downstream area. The proposed desired speed profile is adaptive to traffic states and offers relative robustness, as it only requires one parameter, $w$, to tune.

\subsection{Lower layer: Gap regulation}
While our goal is to manage human drivers' car-following behaviors by proactively increasing the gap, it is crucial to take into account the social acceptance of our approach and ensure that it does not negatively impact traffic throughput. The primary function of the controller's lower layer is to maintain a safe and responsive distance for the ego vehicle, enabling it to adapt to nearby occurrences such as the emergence of large gaps. In particular, when provided traffic state information overestimates or underestimates the \emph{actual} state of traffic, additional feedback mechanisms are employed to allow AVs to respond in a manner more reminiscent of adaptive cruise control (ACC)~\cite{xiao2010comprehensive}. The gap regulation part is designed as ($k_p(h_\alpha - h_\text{des}) + k_d(v_l - v_\alpha)$), which is similar to the design of the ACC vehicle model~\cite{wei2014behavioral}, and aims to maintain a desired time gaps $h_\text{des}$ while further reducing the discrepancy in speeds between ego and leading vehicles. The $k_p$ and $k_d$ terms, similar to other PD feedback control methods, specify the intensity of responses by AVs to such fluctuations. Equipped with the target speed layer, our controller tries to drive smoothly while maintaining a reasonable gap with the anticipation of future oscillations in driving speeds.

\subsection{Safety filter}
Finally, to avoid potential collisions with the preceding vehicle from unforeseen driving events, we additionally restrict the magnitudes of assigned speeds by values imposed by a safety filter $v_\text{fs}$. The safety filter in our formulation is treated as a generic term to allow for flexible assignment. 

In this paper, we adopt a simple method as Eq.~\eqref{eq:safety} with the idea of maintaining sufficiently large gaps, both in space and time, subject to the leading vehicles' most recent fluctuations in speeds.
\begin{equation}
    v_\text{fs} = \frac{s_\alpha - s_\text{min} + v_l\tau_s + \frac{1}{2}a_l\tau_s^2 - \frac{1}{2}v_\alpha\tau_s}{h_\text{min} + \frac{1}{2}\tau_s}, \label{eq:safety}
\end{equation}\vspace{0cm}
where $s$ is the space gap between the preceding vehicle and the subject vehicle; $s_\text{min}$ is the minimum space gap between the two vehicles; $v_l$ is the velocity of the preceding vehicle; $\tau_s$ is the decision-making horizon; $a_l$ is the acceleration of the preceding vehicle; $h_\text{min}$ is the minimum time gap between the two vehicles and $v$ is the velocity of the subject vehicle.

In summary, we develop a two-layer feedback controller designed to:1) proactively and reactively generate a target speed to smooth traffic flow utilizing both macroscopic TSE and microscopic observations; and 2) adjust the gap between vehicles in order to maintain social acceptance and prevent reductions in throughput. o ensure safety, the controller is equipped with a safety filter component. The proposed approach to obtain the control speed $v_c$ is outlined as follows:
\begin{equation}
    v_c = \text{max}(0, \text{min}(v_{\text{target},\alpha} + k_p(h_\alpha - h_\text{des}) + k_d(v_l - v_\alpha), v_\text{fs}) \label{eq:controller}).
\end{equation}

\section{Evaluation in Simulation}

In this section, we evaluate the proposed controller across several simulation experiments. The results aim to answer the following:
\begin{itemize}
    \item Is the proposed controller effective at improving the energy-efficiency and homogeneity of driving across both human-driven and automated vehicles?
    \item Is this approach sensitive to unforeseen events that are common within multi-lane highway networks, and in particular to disturbances induced by sudden and/or aggressive lane changing behaviors?
\end{itemize}

\subsection{Simulation Environment}
To validate the efficacy of our longitudinal driving strategy within I-24, we utilize a microsimulation model presented in~\cite{lichtle2022deploying}. In particular, to capture a degree of variability in driving behaviors that is difficult to recreate with common microsimulation tools~\cite{krajzewicz2012recent, casas2010traffic}, we instead model the platoon response of both simulated human-driven and automated vehicles following leading trajectories collected directly from the target network. These leading trajectories consist of position and velocity measurements $\tau:=\{(x_1, v_1), \ \dots, \ (x_T, v_T)\}$ sampled in increments of $0.1$ second, and vary in terms of time collected and severity of congestion witnessed, thus offering a robust assessment of the influence of automated vehicles within viable states of traffic.


To model the behaviors of platoons of vehicles following the aforementioned trajectories, we initially place $N$ vehicles upstream of the leading vehicle 
and equidistant from one another\footnote{Vehicles are initially placed with $2$-second gaps between one another and driving with the same speed as the leading vehicle.}, and update the state of said vehicles via logic specified either by a car-following model $f_\text{human}(\cdot)$ or the AV model described in the following section. For human-driven vehicles, this acceleration response is dictated by the \emph{Intelligent Driver Model}~\cite{treiber2000congested} (IDM), a popular model for reconstructing string instabilities and the formation of stop-and-go style behaviors. Through this model, the acceleration for a vehicle $\alpha$ is defined by its bumper-to-bumper space gap $s_\alpha$, velocity $v_\alpha$, and relative velocity with the preceding vehicle $\Delta v_\alpha = v_l - v_\alpha$. The fixed parameters for the IDM are set in accordance with~\cite{lichtle2022deploying} and provided in Table~\ref{tab:versions}.





This model is assigned to all vehicles following the leading trajectory when simulating \emph{human-driven} (baseline) responses to varying downstream conditions, while in \emph{mixed-autonomy} simulations, every $\frac{100}{p}$th vehicle is assigned an AV model to mimic a penetration rate of $p\%$.

\begin{table}[]
	\begin{center}
        \caption{Intelligent Driver Model (IDM) Parameters}
	\begin{tabular}{l l l l l l l l}
	\hline
		\textbf{Parameter} & $v_0$ & $T$ & $a$ & $b$ & $\delta$ & $s_0$ & $\epsilon$\\\hline
		\textbf{Value}     & 45 & 1 & 1.3 & 2.0 & 4 & 2 & $\mathcal{N}$(0, 0.3)\\\hline
	\end{tabular}
	\label{tab:versions}
	\end{center}
\end{table}

Figures~\ref{fig:moderate}(a)~and~\ref{fig:heavy}(a) depict the platoon response of human-driven vehicles following a sample of the aforementioned trajectories exhibiting some degree of sharp oscillations in driving behaviors. As seen here, perturbations induced by the leading vehicle are amplified by following vehicles within the platoon 
and propagate backward in space and forwards in time,
resulting in the formation of stop-and-go like behaviors that inhibit the energy-efficiency of the given network.

\subsection{Simulation procedure}

Simulations were conducted on the one-lane environment described above with a step size of $0.1$ sec/step and a following platoon consisting of $200$ vehicles. Among the drives recorded within I-24, we evaluate our method on trajectories that were collected during morning peak demand intervals ($6$am--$7$am) and exhibit some degree of sharp oscillations in driving speeds. This amounts to a total of $10$ varying trajectories, seven of which, we note, observe what may be deemed as \emph{light-to-moderate} congestion (e.g. Figure~\ref{fig:moderate}), whereby vehicles alternate between free-flowing and congested states of traffic, while the remaining three exhibits more \emph{severe} forms of congestion (e.g. Figure~\ref{fig:heavy}), whereby driving speeds are consistently slow and stop-and-go behaviors are frequent. To model an AV penetration rate of $4\%$ we replace every $25$th vehicle model in the platoon with the controller depicted in the previous section. The parameters of this controller used within this assessment are depicted in Table~\ref{tab:controller-params}.

To capture realistic traffic state estimation measurements within the above simulations, we synchronize the above trajectories real world estimates collected from INRIX~\cite{cookson2017inrix}. Historical average speed measurements are collected from INRIX for the target network, and these values are adjusted in position and time with the leading vehicle to produce results similar to those expected in real world settings. These measurements are collected in segments of length approximately equal to $0.5$ miles, and are updated in increments of $1$ minutes.


\begin{table}[]
	\begin{center}
        \caption{Proposed controller parameters.}
	\begin{tabular}{l l l}
    	\toprule
		\textbf{Parameter} & \textbf{Description} & \textbf{Value} \\
		\midrule
		$k_p$ & Proportional gain & $2.0$ \\
		$k_d$ & Differential gain & $0.5$ \\
		$h_\text{des}$ & Desired time headway & $2.0$ s \\
		$w$ &  Sliding window length for speed estimation & $3000$ m \\
		$s_\text{min}$ & Minimum safe space headway & $5.0$ m \\
		$h_\text{min}$ & Minimum safe time headway & $0.5$ s \\
		$\tau_s$ & Safety decision-making horizon & $5.0$ s \\
		\bottomrule
	\end{tabular}
	\label{tab:controller-params}
	\end{center}
\end{table}


\subsection{Performance metrics}
We evaluate the response of vehicles within the above simulation across the following metrics:

\begin{table*}[]
	\begin{center}
        \caption{Simulation Performance}
	\resizebox{0.78\textwidth}{!}{
        \begin{tabular}{l l l l l l}
	\toprule
	& & &
	\textbf{Distance traveled (km)} & 
	\textbf{MPG (AVs)} & 
	\textbf{MPG (total)} \\
	\midrule
	\textbf{Light/Moderate} & &
	Human-driven & 
	$13.71$ & 
    -- & 
	$45.41$ \\
	& \text{Experiment 1} &
	Mixed-autonomy & 
    $13.60$ ($-0.80\%$) & 
	$49.87$ ($+9.82\%$) & 
	$52.76$ ($+16.19\%$) \\
	\midrule
	& &
	Human-driven & 
	$13.86$ & 
    -- & 
	$39.45$ \\
	& \text{Experiment 2} &
	Mixed-autonomy & 
    $13.77$ ($-0.65\%$) & 
	$42.55$ ($+7.86\%$) & 
	$43.48$ ($+10.22\%$) \\
	\midrule
	& &
	Human-driven & 
	$14.58$ & 
    -- & 
	$40.36$ \\
	& \text{Experiment 3} &
	Mixed-autonomy & 
    $14.47$ ($-0.75\%$) & 
	$44.93$ ($+11.32\%$) & 
	$46.66$ ($+15.61\%$) \\
	\midrule
	& &
	Human-driven & 
	$14.09$ & 
    -- & 
	$40.46$ \\
	& \text{Experiment 4} &
	Mixed-autonomy & 
    $14.04$ ($-0.35\%$) & 
	$48.21$ ($+19.15\%$) & 
	$51.17$ ($+26.47\%$) \\
	\midrule
	& &
	Human-driven & 
	$13.23$ & 
    -- & 
	$44.19$ \\
	& \text{Experiment 5} &
	Mixed-autonomy & 
    $13.16$ ($-0.53\%$) & 
	$46.14$ ($+4.41\%$) & 
	$45.11$ ($+2.08\%$) \\
	\midrule
	& &
	Human-driven & 
	$14.24$ & 
    -- & 
	$39.79$ \\
	& \text{Experiment 6} &
	Mixed-autonomy & 
    $14.20$ ($-0.28\%$) & 
	$46.24$ ($+16.21\%$) & 
	$48.16$ ($+21.04\%$) \\
	\midrule
	& &
	Human-driven & 
	$14.48$ & 
    -- & 
	$38.65$ \\
	& Experiment 7 &
	Mixed-autonomy & 
    $14.36$ ($-0.83\%$) & 
	$48.12$ ($+24.50\%$) & 
	$49.31$ ($+27.58\%$) \\
	\midrule
	\textbf{Heavy} & &
	Human-driven & 
	$13.32$ & 
    -- & 
	$36.67$ \\
	& \text{Experiment 8} &
	Mixed-autonomy & 
    $13.31$ ($-0.08\%$) & 
	$42.95$ ($+17.13\%$) & 
	$41.42$ ($+13.50\%$) \\
	\midrule
	& &
	Human-driven & 
	$13.10$ & 
    -- & 
	$36.34$ \\
	& \text{Experiment 9} &
	Mixed-autonomy & 
    $13.00$ ($-0.76\%$) & 
	$52.48$ ($+44.41\%$) & 
	$48.69$ ($+33.98\%$) \\
	\midrule
	& &
	Human-driven & 
	$10.70$ & 
    -- & 
	$30.97$ \\
	& \text{Experiment 10} &
	Control & 
    $10.61$ ($-0.84\%$) & 
	$38.48$ ($+24.25\%$) & 
	$36.03$ ($+16.34\%$) \\
	\midrule
	& &
	Human-driven & 
	$13.53$ & 
    -- & 
	$39.20$ \\
	& \textbf{Average} &
	Mixed-autonomy & 
    $13.45$ ($-0.58\%$) & 
	$46.0$ ($+17.3\%$) & 
	$46.3$ ($+18.0\%$) \\
	\bottomrule
	\end{tabular}}
	\label{tab:performance}
	\end{center}
\end{table*}

\begin{figure}
\includegraphics[width=\linewidth]{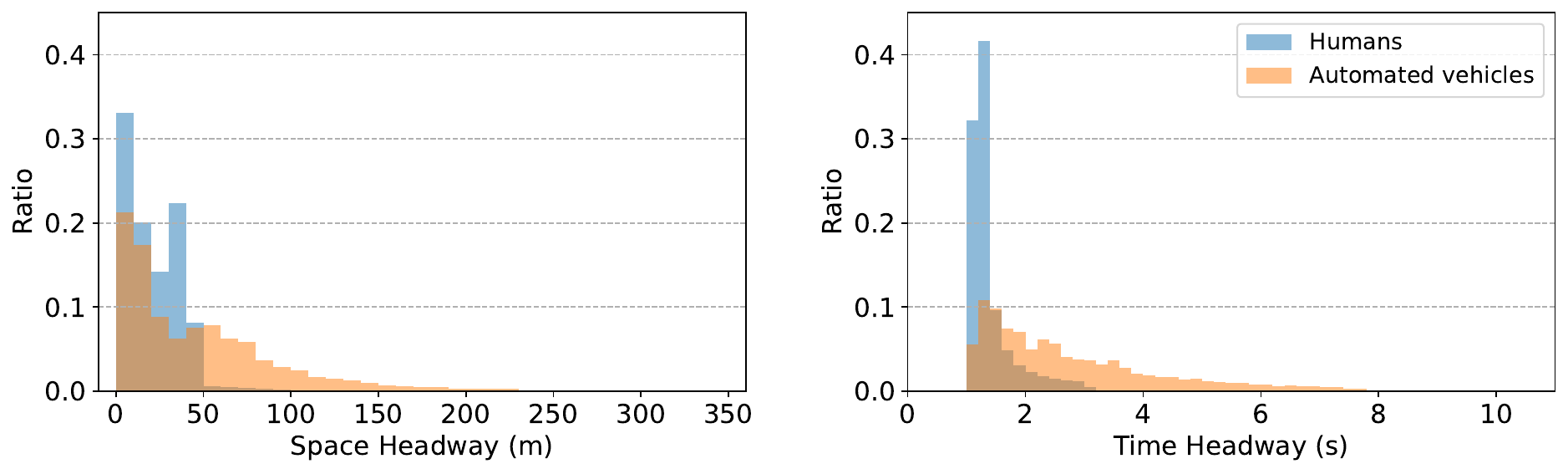}
\caption{Accepted space gap and time gap by human-driven and automated vehicles in each of the fully human-driven and mixed-autonomy simulations. As we can see, automated vehicles primarily maintain reasonable headways but are willing to adopt larger gaps when required to avoid future anticipated congestion.}
\label{fig:headways}
\end{figure}

\begin{enumerate}

\item \textbf{Energy Efficiency.} Improving energy efficiency can incentivize more uniform driving speeds. To analyze the performance of the proposed controller in terms of energy efficiency, we adopt a semi-principled energy model that has a physics-based component~\cite{lee2021integrated}. The model takes as inputs the instantaneous vehicle speed $v$, acceleration $a$, and road grade $\theta$, and outputs engine speed, engine torque, fuel consumption, gear, transmission output speed, wheel force, wheel power, and feasibility of the given ($v$,$a$,$\theta$) with respect to engine speed and engine torque. In our training process, we take Toyota RAV4 as the prototype vehicle and simplify the energy model to a fitted polynomial model with the assumption that the road grade $\theta$ is $0$:
%
\begin{equation}
    g(v,a) = max(f(v,a), \beta)
\end{equation}
where
\begin{equation}
\begin{aligned}
    f(v,a) =& \ C_0 + C_1v + C_2v^2 + C_3v^3 \\
     &+ p_0a + p_1av + p_2av^2 \\
     &+ q_0a_{+}^2 + q_1a_{+}^2v
\end{aligned}
\end{equation}
and $a_{+}$ = max($a$,0), and $\beta$ is the minimum fuel rate, which is not necessarily zero because different vehicles have different criteria for enacting a fuel cut. The parameters we used can be found in Table~\ref{tab:energy-model}.

\begin{table}[]
	\begin{center}
        \caption{The fitted polynomial model Parameters}
        \resizebox{\columnwidth}{!}{%
	\begin{tabular}{l l l l l l}
	\hline
		\textbf{Parameter} & $C_0$ & $C_1$ & $C_2$ & $C_3$ & $p_0$\\\hline
		\textbf{Value}     & 0.14631965 & 0.01217904 & 0 & 0.00002743 & 0.04553801\\\hline
  	    \textbf{Parameter} & $p_1$ & $p_2$ & $q_0$ & $q_1$ & $\beta$\\\hline
		\textbf{Value}     & 0.04743683 & 0.00180224 & 0 & 0.02609037 & 0.01311175\\\hline
	\end{tabular}%
 }
	\label{tab:energy-model}
	\end{center}
\end{table}

The energy consumption obtained from the model will be converted into Miles-Per-Gallon (MPG) as the metric to indicate energy efficiency.

\item \textbf{Throughput.} Since the simulation experiment will end if it reaches the end of the leading trajectory, regulation on controlled vehicles may reduce the throughput near and upstream of these vehicles. For fixed regions, measuring the distance traveled can be an equivalent representation of measuring the traffic flow. Therefore, we use controlled vehicles' travel distance as a representation of the throughput.

\item \textbf{Proximity to leader.} Close proximity may denote unsafe driving behaviors while large distances between vehicles may denote reductions in throughput and may encourage cut-ins and cut-outs by following vehicles. We use space gap and time gap as metrics to measure the proximity to the leader.

\end{enumerate}

\subsection{Comparative analysis}

\begin{figure}
\begin{subfigure}{.24\textwidth}
  \centering
  \includegraphics[width=\linewidth]{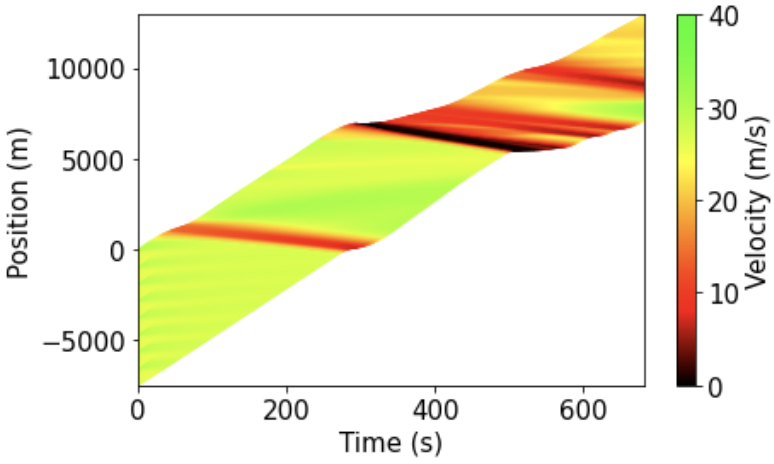}
  \includegraphics[width=\linewidth]{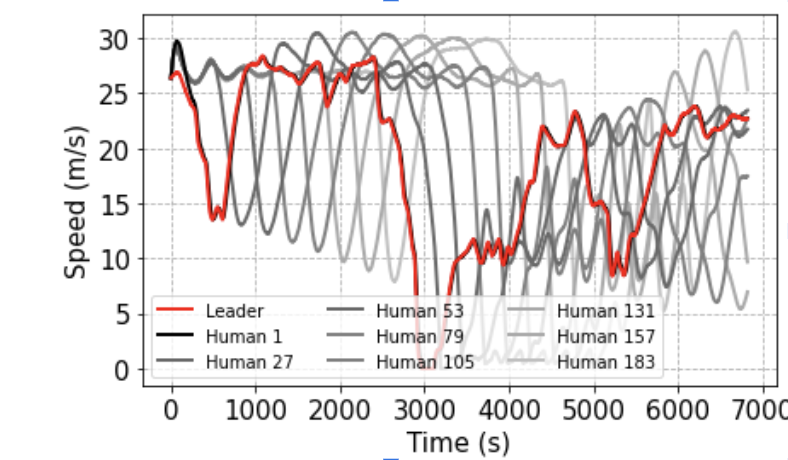}
  \caption{Fully human-driven}
  \label{fig:moderate-baseline}
\end{subfigure}
\hfill
\begin{subfigure}{.24\textwidth}
  \centering
  \includegraphics[width=\linewidth]{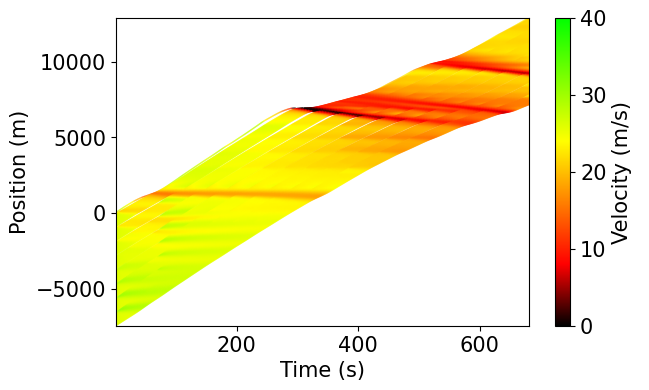}
  \includegraphics[width=\linewidth]{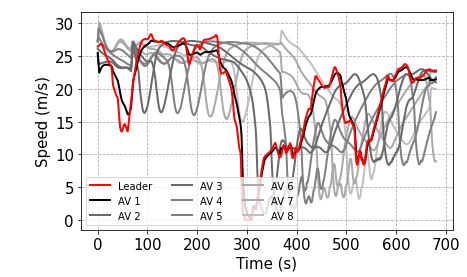}
  \caption{Mixed autonomy}
  \label{fig:moderate-control}
\end{subfigure}
\caption{A sample response from sporadic perturbations induced by a leading trajectory.}
\label{fig:moderate}
\end{figure}
\begin{figure}
\begin{subfigure}{.24\textwidth}
  \centering
  \includegraphics[width=\linewidth]{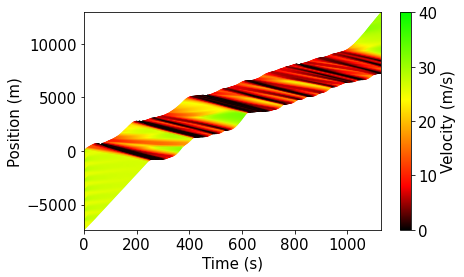}
  \includegraphics[width=\linewidth]{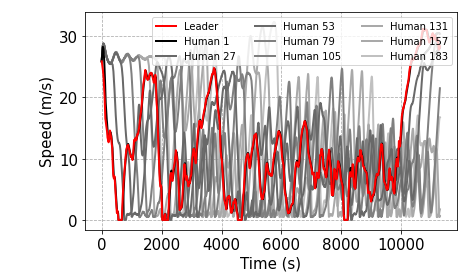}
  \caption{Fully human-driven}
  \label{fig:heavy-baseline}
\end{subfigure}
\hfill
\begin{subfigure}{.24\textwidth}
  \centering
  \includegraphics[width=\linewidth]{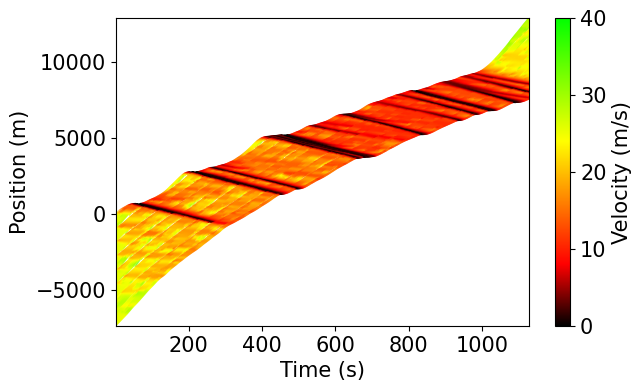}
  \includegraphics[width=\linewidth]{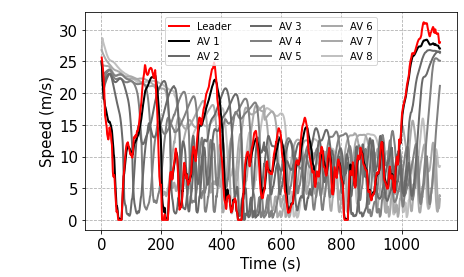}
  \caption{Mixed autonomy}
  \label{fig:heavy-control}
\end{subfigure}
\caption{A sample response from frequent perturbations induced by a leading trajectory.}
\label{fig:heavy}
\end{figure}

Table~\ref{tab:performance} depicts the average performance of the system on all $10$ utilized trajectories for the metrics we described above. Our controller consistently produces driving behaviors that significantly improve the energy-efficiency to both human-driven and automated vehicles at virtually no cost to vehicle miles traveled. Compared to the baseline, the proposed controller provides on average $18.0\%$ savings to energy consumption with only $0.58\%$ reduction in distance traveled by the controlled vehicle.
As Figure~\ref{fig:headways} shows, the controlled vehicles leave a more conservative gap with the preceding vehicle. In controlled cases, both space gap and time gap spread in a wider range. With the knowledge of the congestion downstream, the controlled vehicles should deliberately leave more gaps to avoid sharp deceleration, as a result, drive at a smoother speed and save energy consumption. This behavior propagates to other vehicles immediately upstream of the automated vehicles as well, resulting in more uniform driving speeds throughout the platoon.
This is for instance true in Figure~\ref{fig:moderate}, where perturbations are amplified by trailing drivers within the platoon and result in frequent transitions between free-flowing and congested states of traffic in the Fully human-driven case, while AVs 
dampen the magnitude of oscillations experienced by consecutive vehicles within the platoon as Figure~\ref{fig:moderate}(b) shows. In heavy congestion scenarios represented by frequent perturbations induced by selected leading trajectories, without the guidance of AVs, the strength of severity of oscillations from the leading vehicle produces frequent stop-and-go responses from the upstream, as Figure~\ref{fig:heavy}(a) demonstrates. However, by maintaining speeds near the aggregate state of the network, AVs are capable of negative many of these stop-and-go responses. 

\subsection{Sensitivity to lane changes}

Finally, we evaluate the ability of our approach to cope with external and unforeseen disturbances common to multi-lane networks. In particular, knowing that our method mitigates congestion in part by forming large gaps with leading vehicles when predicted forward speeds are low, we explore the sensitivity of our solution to lane-changing events when AVs form large gaps with their immediate leaders. In order to do so, we use a simple lane change model inspired by the work of~\cite{wu2017multi} that stochastically inserts vehicles into the network when the headway between adjacent vehicles is high and periodically removes vehicles to maintain approximate consistency with the total number of vehicles within a simulation. 

\begin{figure}
\begin{subfigure}{.24\textwidth}
  \centering
  \includegraphics[width=\linewidth]{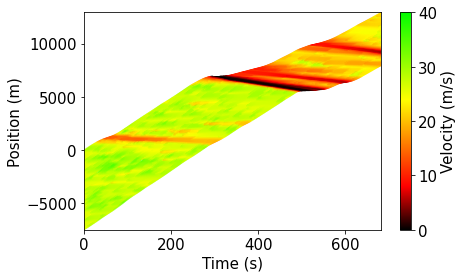}
  \caption{Fully human-driven}
  \label{fig:lc-baseline-tsd}
\end{subfigure}
\hfill
\begin{subfigure}{.24\textwidth}
  \centering
  \includegraphics[width=\linewidth]{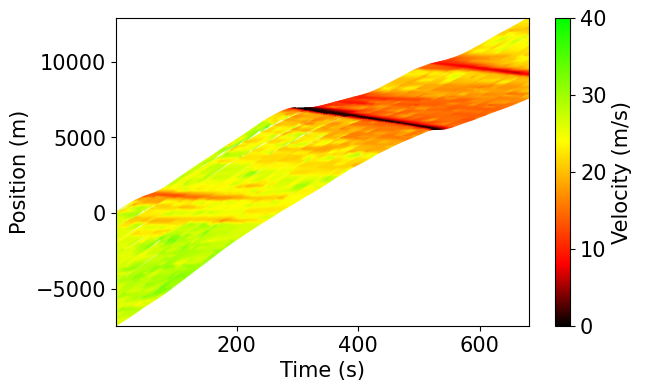}
  \caption{Mixed autonomy}
  \label{fig:lc-control-tsd}
\end{subfigure}
\caption{A sample response from sporadic perturbations induced by a leading trajectory with the simple lane change model.}
\label{fig:lane-change}
\end{figure}

Figure~\ref{fig:lane-change} depicts the spatio-temporal performance of human-driven and automated vehicles when lane changes of the form above are introduced into the simulation environments. As we can see, while more frequent oscillations are observed in the presence of lane changes, AVs continue to produce uniform driving amongst vehicles in the mixed-autonomy settings, providing an on average $9.84\%$ MPG improvement among all 10 experiments. The stochastic injection of vehicles does not result in vehicle-to-vehicle collisions either, demonstrating the effectiveness of the proposed safety filter as well. We leave analyses of this control strategy under more elaborate lane change models for future work.

\section{Application to Field Test}
Motivated by the promising results demonstrated in our simulations, we aim to further validate the modular and flexible architecture of this two-layer control strategy by applying it in a real-world field test.

\subsection{Massive Traffic Experiment}
\begin{figure}
\centering
\includegraphics[height=4cm]{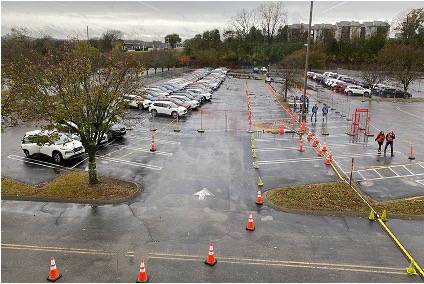}
\caption{The CIRCLES parking lot at capacity after a test, when all vehicles have been returned.}
\label{fig:MVT}
\end{figure}

In mid November 2022, a massive traffic experiment that occurred on the considered network was conducted by the CIRCLES Consortium~\cite{CIRCLES} to test whether introducing just a specific proportion of automated vehicles to the road can help ease the traffic jams and reduce fuel consumption for the traffic system. Over the course of five days, researchers conducted one of the largest traffic experiments of its kind in the world, deploying a fleet of 100 Nissan Rogue, Toyota RAV4 and Cadillac XT5 vehicles onto the considered network during the morning commute. Figure~\ref{fig:MVT} shows the test vehicles in the parking lot at the test site. Each vehicle was equipped with a variety of control algorithms overwriting the cruise control system, and designed to automatically adjust the speed of the vehicle to improve the overall flow of traffic — essentially turning each car into its own “robot traffic manager.”

This massive operation had been prepared over the course of the three last years, parallel to the development of the I-24 MOTION testbed, which equips the considered network with 300 4K digital sensors to monitor traffic. To achieve this tremendous undertaking, more than 50 CIRCLES researchers from around the world gathered in a large “command center” in a converted office space in Antioch, Tenn. Each morning of the experiment, which ran from Nov.14 to Nov.18, trained drivers took the automated vehicles on the designed routes on the I-24 MOTION testbed. As the drivers traversed their route, researchers collected traffic data from both the vehicles and the I-24 MOTION traffic monitoring system. 

Unfortunately, hardware issues arose prior to the commencement of the experiment, preventing us from extracting the preceding vehicles' states as input for the designed controller. As a result, our microscopic observations were limited to the AV's position and velocity. The macroscopic TSE obtained from INRIX remained available to us. In response to this, a ``MegaController" that includes the combination of many researchers' control algorithms in the CIRCLES Consortium was proposed. The two-layer architecture of the proposed method offers the flexibility to adjust and adapt to real-world applications based on different goals and needs. The following section demonstrates how the two-layer controller conceptually adapts to this situation.

\subsection{Application}

\begin{figure}
\centering
\includegraphics[height=4cm]{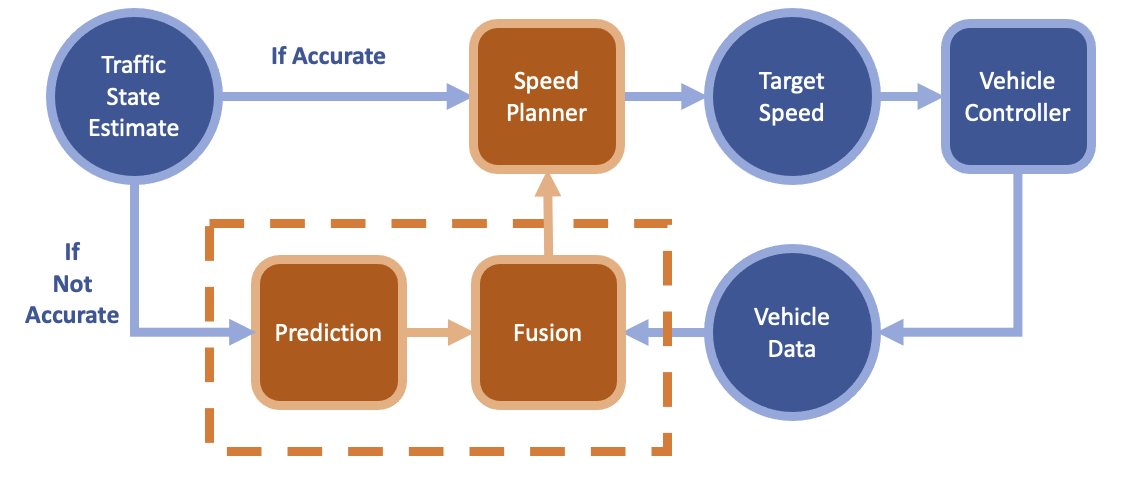}
\caption{The conceptual framework of the proposed controller in application to field test.}
\label{fig:MVT_1}
\end{figure}

The modular architecture of the two-layer controller permits various components to be designed and inserted, depending on the approach and goal. What we present here is a precursor to a more complex architecture that was actually deployed in the field experiment. For additional information on the advanced architecture, please refer to our forthcoming papers, which are currently in preparation. As Figure~\ref{fig:MVT_1} demonstrates, conceptually, the upper layer serves as a ``speed planner" that takes in the downstream traffic state estimates and generates target speeds, while the lower layer can be considered as a ``vehicle controller" that executes the target speeds given the microscopic observations. We had a server on the cloud to compute the target speeds and communicate with each AV. However, traffic state estimates in the real world are usually not accurate enough given the fact of traffic data sparsity and communication delay. For example, in the massive traffic experiment, the INRIX data we utilized was not real-time. The space-mean speed data points received on the server represented the actual traffic state from 3 minutes prior. To compensate for those defects, a module including prediction and fusion was proposed where we would predict the delay of the traffic state estimate for the correction and fuse with the traffic data obtained from our vehicles. This framework aligns with the structure of the ``MegaController" and we incorporated the upper layer of the proposed method into the ``speed planner" module in the field experiment. Due to the massive scale of the experiment, it will take a longer time to mine the data collected, quantify the energy impact and evaluate the methods. We present here a hypothetical reality of the upper layer controller had it been solely run during the field test for demonstration.

\begin{figure}
\centering
\includegraphics[height=3.3cm]{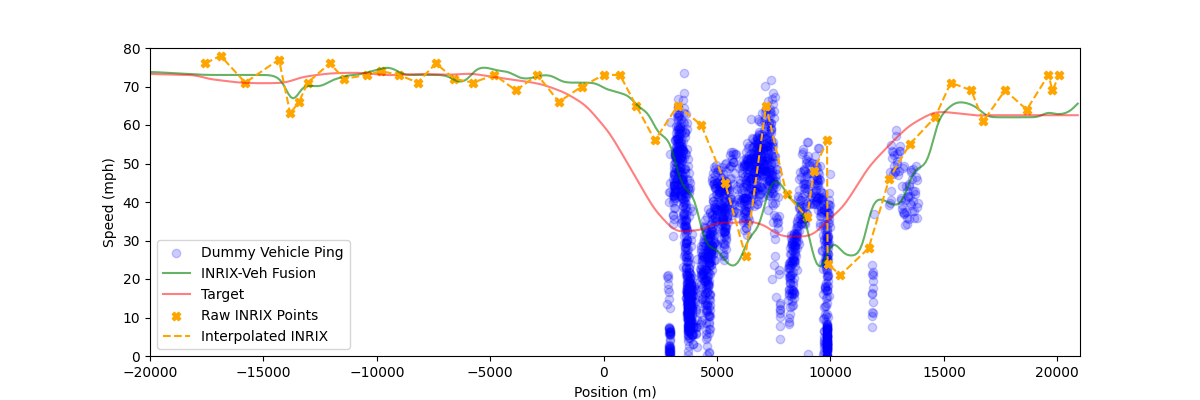}
\caption{Visualization of the procedure of deploying the upper layer on field collected data.}
\label{fig:MVT_2}
\end{figure}

Figure~\ref{fig:MVT_2} illustrates the procedure of deploying the upper layer controller to obtain the target speed profile based on real traffic state estimates provided by INRIX during the course of our field test. The raw data to begin with, shown as the orange crossings, is the average speeds across multiple pre-defined segments on the test routes in one minute. We simulate the vehicle ping records that contain 31 dummy vehicles' positions and speeds at 1 Hz.  
The green curve in the figure is the traffic state estimation obtained by fusing vehicle ping data (blue dots) with interpolated INRIX data (dashed orange line). Feeding it into the speed planner, we can obtain the target speed profile shown by the red curve. Obviously, compared to the other curves in the figure, the target profile can guide vehicles to slow down earlier and pass through the congested area at a relatively smooth speed, thus aiming to achieve harmonization of the overall traffic flow.

\section{Conclusion}

This paper explores the problem of designing congestion mitigation control strategies through automated vehicles. We depict a two-layer control strategy that utilizes downstream traffic state information to plan and coordinate a smoother driving trajectory for the purpose of harmonizing driving speeds and improving energy efficiency. Evaluated with simulations that capture a high degree of variability in driving behaviors and traffic state estimates common to real-world networks, our proposed method could achieve an average of over $15\%$ energy savings with only $4\%$ AVs introduced to the simulated Interstate I-24 network.
We demonstrate the proposed method's modular and flexible architecture with its application to the massive traffic experiment.
Future works can include extending this with more accurate and robust simulations of traffic flow dynamics, and devising methods for performing similar congestion mitigation without the need for downstream traffic state estimates.

\section*{Acknowledgements}
This material is based upon work supported by the National Science Foundation under Grants CNS-1837244/CNS-2135579. This material is based upon work supported by the U.S.\ Department of Energy’s Office of Energy Efficiency and Renewable Energy (EERE) under the Vehicle Technologies Office award number CID DE--EE0008872. The views expressed herein do not necessarily represent the views of the U.S.\ Department of Energy or the United States Government.

\bibliographystyle{IEEEtran}
\bibliography{reference}
\end{document}